# Two Dimensional Crystal Tunneling Devices for THz Operation

B. D. Kong[1], C. Zeng[2], D. K. Gaskill[3], K. L. Wang[2], and K. W. Kim[1*]

[1]Department of Electrical and Computer Engineering, North Carolina State University, Raleigh, North Carolina, 27695-7911, United States of America

[2]Department of Electrical Engineering, University of California – Los Angeles, Los Angeles, California, 90095, United States of America

[3]U.S. Naval Research Laboratory, 4555 Overlook Avenue, Washington, D.C., 20375, United States of America

[*]Electronic address:kwk@ncsu.edu

## ABSTRACT

Two dimensional (2D) crystal heterostructures are shown to possess a unique opportunity for novel THz nonlinear devices. In contrast to the oxide tunneling barrier, the uniformity of 2D insulators in the thickness control provides an ideal condition for tunneling barriers in the atomic scale. Numerical calculations based on a first principles method clearly indicate the feasibility of diode operation with barriers as thin as two monolayers of hexagonal boron nitride or molybdenum disulfide when placed between graphene-metal asymmetric electrodes. Further analysis predicts the cut-off frequencies of the proposed device over 10 THz while maintaining strong nonlinearity for zero-bias rectification. Application of the tunneling structure to hot electron transistors is also investigated, illustrating the THz operation with superior power performance. The proposed concept provides an excellent opportunity for realizing active nonlinear devices in the frequency range inaccessible thus far.

The success of graphene has sparked significant interest in the potential advantages of two dimensional (2D) crystals in general. This class of materials, also known as van der Waals crystals, offers a unique opportunity for realizing atomic scale structures, including ultra-thin multi-layer structures of unusual material combination, with the potential for uniformity and less sensitivity to process variations. With recent experimental confirmation of honey-comb lattice silicene,[1] 2D Crystals now cover a broad range; i.e., semimetals, semiconductors, and insulators – graphene, silicene, molybdenum disulfide, hexagonal boron nitride (*h*-BN), etc. Along with the investigations on fundamental materials properties, much of the attention has been focused on in-plane carrier transport within the framework of field-effect transistors.[2,3]

In this letter, we examine the properties of 2D crystal tunneling structure and their application to very high frequency operation that has so far been inaccessible by conventional materials. The structure utilizes *h*-BN or $MoS_2$ as an ideal 2D insulating barrier with atomistic uniformity in the thickness control, while graphene plays the role of an asymmetric contact along with a metallic electrode. Tunneling applications usually pursue two advantages; ultra-short tunneling time[4] for high speed operation and highly nonlinear resonant tunneling for switching and oscillator circuits. While the latter highlights the necessity of quantized energy states such as quantum well, the nonlinearity can also be realized by simply combining dissimilar metal electrodes at a single tunneling barrier. This essentially constitutes a tunnel diode akin to the metal-insulator-metal (MIM) structure previously proposed[5] without the detriment of nonuniform oxide found in early investigations.[6] Despite the promise for THz operation utilizing the electron transit time on the femtosecond order, it is well known that thin metal-oxide films cannot maintain the thickness in the angstrom scale uniformly over a sizable area, leading to device-to-device performance variation.[6] Even a small deviation of oxide thickness could lead to current crowding at the thinnest point and subsequently to drastic reduction in the effective area of the device, explaining at least in part the limited advances of the MIM tunnel structure. In this regard, 2D insulating or semiconducting crystals are ideal candidates for the tunneling barriers. Specifically *h*-BN is a wide bandgap material whose (bulk) gap energy has been measured in the range of approximately 3.8 – 5.8 eV,[7] whereas $MoS_2$ shows an intermediate value with ~1.3 – 1.9 eV.[8] This choice is expected to provide a criterion for the selection of 2D crystal as a tunnel insulator. Further, it also enables consideration of a multi-layered structure consisting of two or more insulating and semiconducting 2D crystals to enhance functional nonlinearity such as amplification. As for the electrodes, Cu is our choice to be paired against graphene. The Cu(111) surface has a lattice constant that almost matches to graphene and *h*-BN; in fact, *h*-BN and graphene growth on catalytic metal surfaces has a long and relatively successful history (at the time graphene was called monolayer graphite).[9,10] In addition, successful formation of layered $MoS_2$ structure has been reported.[11,12] Accordingly, no major technical difficulties are expected in the fabrication of the desired graphene/2D insulator/metal (GIM) structure; i.e., G/BN/Cu or G/$MoS_2$/Cu. Moreover, substantial difference in the work functions of Cu(111) and graphene (4.94 eV vs. 4.2 – 4.6 eV depending on the number of graphene layers)[13] naturally ensures an asymmetric band profile that eliminates the need for a dc bias for ac rectification unlike in the symmetric MIM device.

Analysis of the proposed structures requires a computational approach that can handle the atomic details including potential formation of interface states. It is well known that the presence of heterogeneous boundaries with nanometric separation could alter the material properties from those of the bulk. We apply density functional theory (DFT) to obtain the electronic structures of the GIM device accurately. Then, the calculation of carrier transport and subsequent current-voltage (I-V) characteristics are approached by the non-equilibrium Green's function method based on the electronic wave functions attained from DFT. Siesta and Transiesta [14] packages are used for this purpose with the pseudopotentials in the local density approximation. Characterization of the Cu(111) slab illustrates the accuracy of the DFT based treatment; the obtained work function of 4.99 eV is close to the experimental value 4.94 eV. [15] As for graphene, the work function varies with the number of layers. [13] Our calculation predicts 4.33 eV for the 6-layer A-B stacked graphene slab, which is also in close agreement with experimentally observed 4.55 eV. [13] In these two contact materials, the work function is a crucial parameter for the desired asymmetric characteristics. As for the insulating materials, 4.28 eV is obtained for the bulk $h$-BN gap in concert with previous studies. [7, 16] A direct comparison for thin films of atomic thickness is rather difficult as the experimental values vary widely. [17] For $MoS_2$, our result of 1.87 eV (direct) and 1.15 eV (indirect) also compares favorably with very recent measurement of approximately 1.9 eV (direct) and 1.6 eV (indirect) for the single- and double-layer cases, respectively. [8]

All of the considered heterostructure geometries are optimized by minimizing the interatomic forces. For the case with $h$-BN barrier, two electrodes are constructed with 9 Cu layers and 6 graphene layers, respectively, while the insulator is varied between 2 to 4 monolayers in thickness. Since the most likely scenario is that graphene and $h$-BN are either grown directly or transferred on a Cu(111) substrate, the lattice constant is chosen to match the bulk value of this material (2.51 Å) resulting in negligible strain (0.01 %) on graphene or $h$-BN. From the Cu substrate to $h$-BN, the atoms are arranged in the most stable phase found from earlier studies. [18] Specifically, the first $h$-BN layer has N placed on the same site of the Cu atoms and B on the empty fcc site. Subsequent layers of $h$-BN are stacked in AB sequence as in graphene. The numbering convention used for the insulating layers starts from the Cu(111) surface. Unlike the tunnel barriers, the exact number of electrode layers chosen for the calculation is not crucial in the device performance. For instance, a couple of monolayers (instead of 6 as specified earlier) will also work similarly for graphene.

Calculated electronic structures of the GIM system reveal that the electronic properties in each material (i.e., G, I, and M) deviate from those of the isolated counterpart. Figure 1 shows the electronic energy dispersion and the projected density of states (PDOS) when two monolayers of $h$-BN is used in the GIM structure (i.e., 2-BN). To distinguish the changes in the energy bands, the projected bulk band structures of graphite and Cu(111) are also plotted by the green and orange shaded regions, respectively. The result indicates that the band structure of the insulating layer is substantially changed after formation of the GIM structure. The modification is particularly pronounced in the first $h$-BN layer where the conduction band edge experiences a downward shift near the K point in the Brillouin zone; see the arrow marked A in Figure 1(a). Accordingly, the conduction band minimum of the first layer is now much lower than that of the second layer which is marked B. This is also confirmed by the PDOS plot in Figure 1(b). As shown, the data clearly indicate formation of interface states within the gap (between -2 eV and 2 eV) although the magnitude is relatively small. For the second $h$-BN layer, the influence of Cu in contact becomes much less significant and the conduction band approaches to the normal value. The electronic structures of GIM with different numbers of $h$-BN layers show a similar trend. Hence, an intact potential barrier is formed from the second layer and beyond. Contrary to the metal/BN interface, the BN/G interface seems relatively inert. However, this is actually the case only for the $h$-BN side. The effect of interface on graphene is discussed later.

For the MoS$_2$ GIM, we use the $\sqrt{3} \times \sqrt{3}\ R\ 30°$ symmetry for the insulating layer to match the lattice structure to Cu(111). According to recent experimental data,[12] single-layer MoS$_2$ on graphite is rotated 5° with respect to the void site of the substrate. Since the lattice constant of Cu(111) is almost the same as that of graphite, it is probable that MoS$_2$ would assume a similar interface structure with Cu(111), in which case the necessary supercell size is very large. The lattice constant of unstrained MoS$_2$ is 3.15 Å,[12] whereas it becomes 2.9 Å in the studied $\sqrt{3} \times \sqrt{3}\ R\ 30°$ structure – a strain of 7.7 %. Nevertheless, the impact of this simplification on the calculated bandgap is relatively minor at about 0.1 eV when the thickness of MoS$_2$ is two monolayers or larger. In the $\sqrt{3} \times \sqrt{3}\ R\ 30°$ structure, two different atomic arrangements are possible. One is the case where the S atoms are placed on top of Cu atoms and the other is with all S atoms on empty sites. After geometry optimization, the difference in the total energy of two configurations is negligibly small at 0.08 eV. Since both configurations give almost the same I-V characteristics, only the results from the first case are presented. The PDOS for each layer of double-layer MoS$_2$ (i.e., 2-MoS$_2$) is shown in Figure 2, where the electronic structure follows a trend very similar to the insulating barrier of h-BN GIM. The band structure of the first MoS$_2$ layer is modified and the main potential barrier appears from the second layer. Incidentally, the indirect bandgap of 1.53 eV extracted from the second layer corresponds to that of 2-MoS$_2$ (~1.6 eV). One difference, when compared to h-BN, is that metal-induced gap states are more pronounced in this case, indicating stronger interaction between Cu and MoS$_2$ at the interface.

This difference between the h-BN and MoS$_2$ cases are more clearly illustrated in Figure 3, where charge re-distribution is calculated by $\Delta\rho = \rho_{GIM} - \rho_G - \rho_I - \rho_M$. Note that $\rho$ is the electron charge density, and the red and blue colors represent iso-density surfaces with positive and negative $\Delta\rho$, respectively. At the BN/Cu interface shown in Figure 3(a), a dipole formation is clearly visible as the Cu side is charged more negatively and the h-BN more positively. This is nothing but a case of Schottky barrier lowering which is confirmed by the energy band plot in Figure 1(a); per our discussion earlier, band A that belongs to the first h-BN layer is substantially lowered when compared to its bulk position. On the other hand, charge re-distribution occurs more vigorously at the MoS$_2$/Cu interface in Figure 3(b).[19] In addition, introduction of pronounced interfacial states below Fermi energy level (E≤0 in Figure 2) indicates that these states are filled in equilibrium. Consequently, it appears that a weakly hybridized bonding exists between Cu and S atoms. Our calculation of the binding energy also suggests 0.32 eV and 0.89 eV per unit cell at the BN/Cu and MoS$_2$/Cu surface, respectively. Hence, judging from the conventionally accepted criteria,[20] MoS$_2$ is chemisorbed to the Cu(111) surface in the studied configuration whereas h-BN is considered physisorbed. In both cases, it appears that more than one monolayer is needed to function as an effective tunnel barrier. Clearly, the influence of a metallic material in contact cannot be ignored at the interface.

Another interesting point is the insulator/graphene junction. As indicated in Figure 3(a), carbon atoms at the interface with h-BN are polarized. It can be easily expected that the polarization breaks the symmetry of graphene π-orbitals and opens a small bandgap. Actually this effect can be seen in Figure 1(b) from a pronounced peak of the graphene PDOS near the Fermi level $E_F$ (see the black arrow in the last panel on the right). In ideal graphene, the DOS goes to zero at $E_F$. The peak of DOS means that there is a band edge since the DOS is proportional to the inverse of $|\nabla_k E(k)|$. If a gap exists, there should be two peaks. In the present case, the resulting value (18 meV) is simply too small to be discerned from the scale of the figure. A similar polarization effect is also observed at the MoS$_2$/G interface but the relative magnitude of charge re-distribution is smaller than MoS$_2$/Cu, making it practically unnoticeable in Figure 3(b).[19] Nevertheless, it is clear from the results that the electronic properties of the first-layer graphene at the interface deviate from the ideal case. For instance, the in-plane electric conductivity is expected to be lower due to the increased scattering.

The results thus far demonstrate that the details of atomic potential and the interfacial states affect a large portion of the tunneling barrier shape, clearly revealing the severe limitation of the traditional trapezoidal potential barrier approximation when the thickness of the insulating layer shrinks to atomistic scale. Accordingly, electron transmissions in the GIM structures must also be described to maintain the accuracy of atomic modeling. Considering that the transport is mostly ballistic in this study, we adopt the Green's function method based on the DFT Hamiltonian and the Landauer-Büttiker formalism. A 12 ×12 $k$-point mesh is used for the $k_x$-$k_y$ plane ($k_z$ is the parallel to the current) and the results are tested with denser meshes up to 60 ×60 to ensure the numerical stability. The selected results are shown in Figure 4(a) and (b). The nonlinearity is estimated by fitting the I-V curves to a third-order polynomial $I = R_D^{-1}(V + mV^2 + nV^3)$,[21] shown by the red dashed lines in the figure while the extracted parameters are summarized in Table 1. As the GIM structure exhibits a barrier thickness dependence that is qualitatively similar to the classical cases,[21] two competing requirements become apparent. Namely, a thinner barrier is desirable from the resistance aspect but not so for the nonlinearity. Nevertheless, even the thinnest case (i.e., two monolayers of $h$-BN) provides substantial nonlinearity, indicating rectification as a potential application. Non-zero $m$ represents ac rectification at zero bias. More generally, the calculated curvature coefficient $\gamma$ ($= \frac{\partial^2 I}{\partial V^2} / \frac{\partial I}{\partial V}$)[22] in Figure 4(c) and (d) further confirms the nonlinear I-V characteristics of GIM. Use of MoS$_2$ in place of $h$-BN is advantageous for both rectification and detection. At two monolayers (MoS$_2$), the actual thickness corresponds to that of 3 $h$-BN layers. Combined with the low barrier height, MoS$_2$ can achieve low tunnel resistance and large nonlinearity simultaneously. This suggests that yet another 2D crystal with a smaller bandgap (such as silicene with a buckled lattice grown on Ag)[1] could further improve the characteristics as a larger insulator thickness may become acceptable while maintaining the tunnel resistance low.

The obtained nonlinearity of the proposed GIM structure compares favorably to corresponding MIM diodes. In a recent measurement,[22] the zero-bias curvature coefficients of Al/AlO$_x$/Al, Al/AlO$_x$/Ti, Al/AlO$_x$/Ni, and Al/AlO$_x$/Pt air-oxidized structures were estimated to be 0.00, 0.08, 0.45, and 0.74 V$^{-1}$, respectively. In comparison, the proposed GIM with MoS$_2$ barrier provides a substantially larger value of approximately 1.5 V$^{-1}$. The $h$-BN GIM also shows comparable nonlinearity whose $\gamma$ at zero bias is 0.75 V$^{-1}$. Actually, in the metal-oxide MIM case, the undesirable irregularity of air-oxidation may be at least partly responsible for the nonlinearity as the structures with controlled oxidation (for uniformity) apparently experience reduction in $\gamma$.[22] Thus, the advantages of 2D crystals are expected to be even more obvious when the uniformity requirement is ensured.

It is interesting to compare our calculated I-V characteristics with relevant experimental data in the literature. Very recently, a tunneling field-effect transistor with a G/BN/G structure was examined,[23] where the extracted tunneling resistance was 100 GΩ/μm$^2$ for 6 ± 1 layers of $h$-BN barrier at a low back gate voltage. If our results are extrapolated to 5 – 6 layers via a classical model, the value would be somewhat larger being in the range 300 – 900 GΩ/μm$^2$. The discrepancy can come from a number of contributions. For one, the thickness variation in the experimental device would provide a reduced effective resistance as the current flows through primarily the thinnest part. In addition, surface microstructures which may depend on the fabrication method could influence the interface states (thus, the tunneling currents) as shown earlier. Furthermore, the limited predictability of the theoretical calculation on material properties (such as the band alignment, gap, etc.) inevitably impacts the outcome. Other tunneling mechanisms such as phonon assisted tunneling can also contribute to the difference. Nevertheless, the relative closeness of the two results illustrates the accuracy of our calculation.

Since the tunneling time is fast (~ $10^{-15}$ s), the GIM structure can maintain dc characteristics up to at least several hundred THz intrinsically.[4] However, the frequency response is also limited by the parasitic RC time constant of the connected circuit network. Since an immediate application of GIM is a rectifier in the IR frequency range, the operation frequency of GIM structures can be estimated by assuming that it is tied to an ideal half-wave thin dipole antenna with impedance of 73.1 Ω.[21] One uncertainty is the capacitance of the insulating layers, particularly the dynamic response of aforementioned interfacial states of the first layer. In addition, the relative permittivity could change when the insulating barrier becomes atomistic in scale. We assume the possible maximum capacitance (i.e., the worst case) by treating the first barrier layer as a part of the electrode along with the bulk permittivity values (parallel to the *c* axis); i.e., 4.53 for *h*-BN [24] and 2.74 for $MoS_2$.[25] For *h*-BN, $f_C$ can go above 10 THz when the area becomes less than approximately 60 nm× 60 nm. The same frequency range is accessible by $MoS_2$ with a larger cross-section of around 90 nm× 90 nm.

There is an additional opportunity to improve the nonlinearity of GIM by combining two dielectric materials with different potential barrier heights.[26] To this end, a combination of *h*-BN and $MoS_2$ is also examined by using the same approach; namely, the graphene-insulator-insulator-metal (GIIM) structure. Specifically, the I-V characteristics of two GIIM structures, G/1-BN/2-$MoS_2$/Cu and G/1-$MoS_2$/2-BN/Cu (i.e., two single-layer/double-layer combinations of *h*-BN and $MoS_2$), are analyzed as shown in Figure 5(a) and (b), respectively. The results are now very nonlinear; nearly exponential similar to the responses of pn diodes to the point that it is not possible to fit a polynomial function any longer. The differential resistance $\partial V/\partial I$ at V=1.5 V is estimated to be 99 Ω/μm$^2$ for G/1-BN/2-$MoS_2$/Cu and 211 Ω/μm$^2$ for G/1-$MoS_2$/2-BN/Cu, respectively, which are substantially smaller than the low-bias resistance that is comparable to $R_D$ of GIM. The frequency characteristics of the GIIM structures can be estimated in the same manner as in the GIM. Due to the increased thickness, the capacitance decreases and the same $f_C$ can be achieved with a wider device area. For instance, the 10-THz operation is accessible with the area of 100 nm×100 nm for both structures. Moreover, the current drive is fairly comparable to the $MoS_2$ GIM.

The diode operation discussed above can be readily extended to a tunnel transistor. Two reversely connected diodes form a transistor, which can be imagined by the Ebers-Moll model. In this case, however, the injected hot carrier plays the role of minority carrier in bipolar junction transistors. The gain of this hot electron transistors (HET) can be improved by utilizing a Schottky contact on one side of the device.[27] With recent demonstration of graphene-semiconductor Schottky junction,[28] no major difficulty is expected in the realization of the HET based on the proposed GIM structure. Forming a few layers of 2D insulator on a metal surface is a well-established technology and graphene layers can be grown on SiC epitaxially.

Accordingly, transplanting a metal-insulator structure on epitaxially grown graphene is a clearly realizable strategy. One such example is illustrated in Figure 5(c), where a GIM diode constitutes the emitter–base junction and the graphene/semiconductor interface forms the base–collector junction. Due to the short transit time, the device switching is essentially dominated by the RC charging time; i.e., $\Delta t \sim R_B(C_{BE} + C_{CB})$, where $R_B$, $C_{BE}$, and $C_{CB}$ are the base resistance, emitter-base capacitance, and base-collector capacitance, respectively.[27] In a forward bias case, $C_{CB}$ can be ignored and $\Delta t$ is further simplified to $R_B C_{EB}$. Assuming 150 Ω for the graphene sheet resistance,[29] the required device dimensions for $f_C$ of 1 THz and 5 THz are 200 nm×200 nm and 90 nm×90 nm, respectively, with the double-layer MoS$_2$ insulating barrier. From the I-V data in Figure 4, it is estimated that these devices can provide approximately 0.75 mW and 0.15 mW with 3 V applied bias (V$_{BE}$ = 1V and V$_{CB}$ = 2V) at the frequencies below $f_C$ (i.e., 1 THz and 5 THz), respectively. When double-layer h-BN is used, the same $f_C$ can be achieved with similar device dimensions of 140 nm×140 nm (1 THz) and 60 nm×60 nm (5 THz), but the corresponding power is only in the nW range. However, it is important to note that the actual power level could be substantially higher than the obtained numbers since the present calculation may have overestimated the tunneling resistance (thus, underestimating the current) than the measured values as discussed earlier. Furthermore, use of another 2D crystal with a lower barrier height could drastically improve the current/power characteristics as discussed earlier (e.g., the Ag/silicene/G combination). Nonetheless, these numbers (particularly, with MoS$_2$) are very encouraging as the frequency domain near 10 THz has been unattainable with the conventional devices. The fastest transistor reported thus far is an InP pseudomorphic heterojunction bipolar transistor (HBT) whose $f_C$ is estimated around 710 GHz.[30] Considering that the pseudomorphic HBT and the GIM HET share similar technical challenges, the structural simplicity of the GIM HET offers a significant advantage by circumventing many of the potential issues in nanoscale device fabrication such as doping. At the same time, the ballistic nature of the GIM HET (thus, less noise) could prove to be beneficial in noise-sensitive devices [31] along with superior speed. The proposed GIM HET is expected to make a major impact on a number of applications including low-noise amplifier, pre-amplifier, and voltage controlled oscillator with operating frequencies over a few THz that has so far been inaccessible.

In summary, an atomically thin tunneling device is proposed by taking advantage of 2D crystals and asymmetric metal electrodes. Using the first principles approach, the detailed nature of heterogeneous metal/insulator and graphene/insulator interfaces is analyzed and the potential applications examined. The calculated I-V characteristics evidence the nonlinear operation of the proposed GIM and GIIM structures, clearly demonstrating the feasibility as a zero-bias rectifier in the frequencies over 10 THz. It is also shown that the GIM and GIIM structures can be readily extended to a transistor for THz amplification by coupling the tunneling diode with a graphene-semiconductor Schottky contact (or, alternatively, another 2D tunnel barrier). The predicted THz operations and possibilities of further improvements via judicious choice of materials and device parameters clearly illustrate that the significant potential impact of the proposed concept for realizing active nonlinear devices in a frequency range that is beyond the reaches of conventional semiconductor counterparts.

This work is supported, in part, by the SRC Focus Center on Functional Engineered Nano Architectonics (FENA), US Army Research Office, and the Office of Naval Research.

**Table 1.** Calculated resistance and nonlinearity of the proposed GIM structures

| Layer Num. | $R_D$ ($\Omega/\mu m^2$) | m | n |
|---|---|---|---|
| 2 $h$-BN | $5.36 \times 10^7$ | 0.379 | 0.657 |
| 3 $h$-BN | $1.18 \times 10^9$ | 0.710 | 0.801 |
| 4 $h$-BN | $2.26 \times 10^{10}$ | 1.18 | 1.19 |
| 2 MoS2 | $6.58 \times 10^2$ | 0.757 | 1.89 |

Figure Captions

**Figure 1.** (Color online) (a) Energy band diagram of 2-BN GIM (i.e., with two layers of $h$-BN). The green and orange shaded regions are the projected bulk band structure of graphene and Cu(111). Points marked by A and B denote the conduction band of the first and second $h$-BN layer, respectively. (b) PDOS of interfacial atoms. For the Cu electrode, the line represents the PDOS of the Cu atom at the surface. In the case of h-BN layers, the line represents the total PDOS of B and N atoms. The energy gap region with zero PDOS corresponds to a potential barrier in the classical plain wave tunneling approximation. As for the graphene side (G), the line denotes the total PDOS of two carbon atoms in the unit cell. The arrow highlights a pronounced peak near $E_F$. The energy level is adjusted with respect to the Fermi energy (i,e,, $E_F = 0$).

**Figure 2.** (Color online) PDOS of 2-MoS$_2$ GIM. The Cu electrode is placed on the left side and the graphene (G) is on the right. As in Figure 1, the energy level is adjusted with respect to the Fermi energy.

**Figure 3.** (Color online) Difference in the charge density after the GIM structure is constructed with two monolayers of (a) BN or (b) MoS$_2$. The red color indicates the increased electron density (negative), while the blue symbolizes the opposite (positive). Essentially, the plot reveals how the electrons are shifted from their neutral positions to form the interfaces. (a) and (b) show the iso-density surface $\Delta\rho = 0.001$ and $\Delta\rho = 0.002$ (in absolute units), respectively. [19]

**Figure 4.** (Color online) I-V characteristics of (a) 2-BN and (b) 2-MoS$_2$ GIM. The red dashed lines represent 3rd order polynomial fits. The extracted curvature coefficient γ of 2-BN and 2-MoS$_2$ is plotted in (c) and (d), respectively.

**Figure 5.** I-V characteristics of (a) G/1-BN/2-MoS$_2$/Cu and (b) G/1-MoS$_2$/2-BN/Cu GIIM. (c) Schematic illustration of a transistor utilizing a 2D crystal system. The envisioned device resembles a hot-carrier transistor. Abbreviation E, B, and C stand for emitter, base, and collector. The base-collector junction can also be formed by a combination of two 2D crystals instead of graphene/semiconductor as shown.

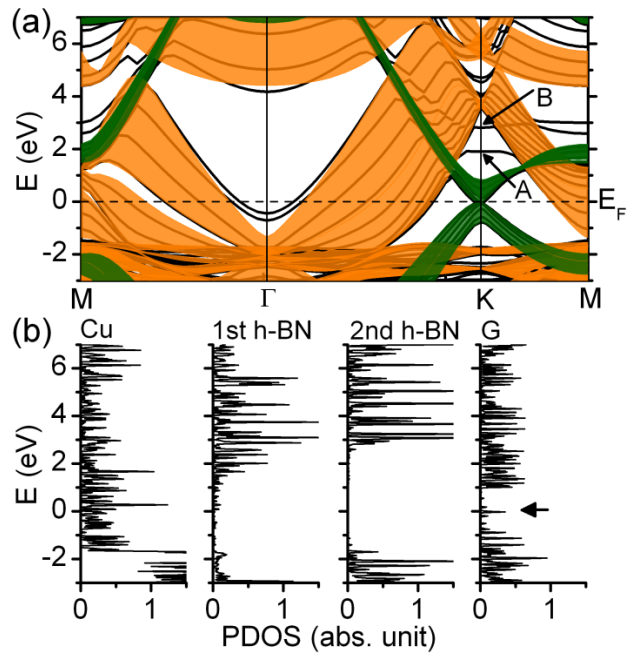

**FIG 1:** Kong et al.

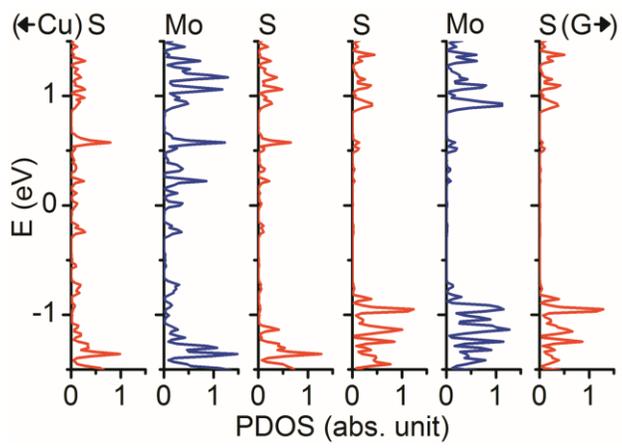

**FIG 2:** Kong et al.

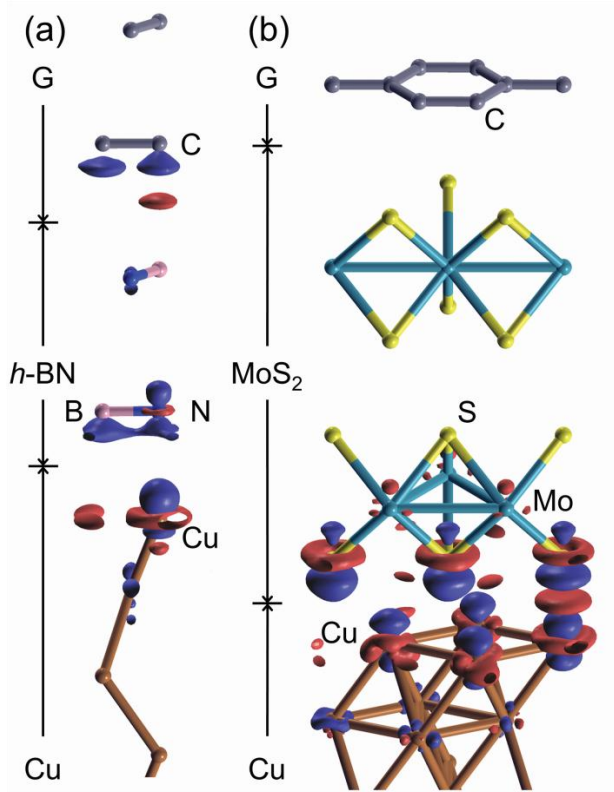

**FIG 3:** Kong et al.

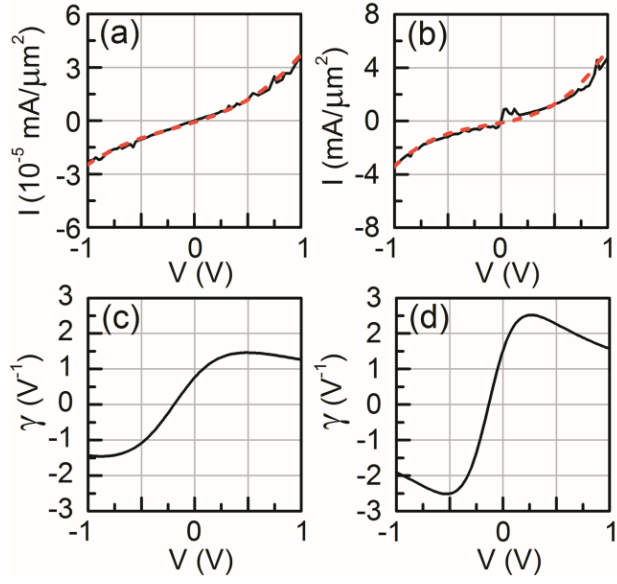

**FIG 4:** Kong et al.

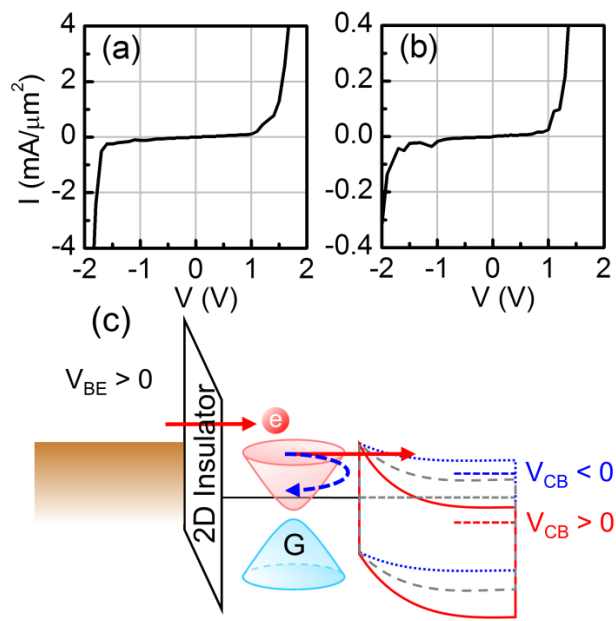

**FIG 5:** Kong et al.